# Boron Nitride Nanosheet Veiled Gold Nanoparticles for Surface Enhanced Raman Scattering

*Qiran Cai,[1] Srikanth Mateti,[1] Kenji Watanabe,[2] Takashi Taniguchi,[2] Shaoming Huang,[3]*

*Ying Chen[1]\* and Lu Hua Li[1]\**

1. Institute for Frontier Materials, Deakin University, Geelong Waurn Ponds Campus, VIC 3216, Australia

2. National Institute for Materials Science, Namiki 1-1, Tsukuba, Ibaraki 305-0044, Japan

3. Nanomaterials and Chemistry Key Laboratory, Wenzhou University, Wenzhou 325027, China.

## ABSTRACT

Atomically thin boron nitride (BN) nanosheets have many properties desirable for surface enhanced Raman spectroscopy (SERS). BN nanosheets have a strong surface adsorption capability towards airborne hydrocarbon and aromatic molecules. For maximized adsorption area and hence SERS sensitivity, atomically thin BN nanosheet covered gold nanoparticles have been prepared for the first time. When placed on top of metal nanoparticles, atomically thin BN nanosheets closely follow their contours so that the plasmonic hot spots are retained. Electrically insulating BN nanosheets also act as a barrier layer to eliminate metal-induced disturbance in SERS. Moreover, the SERS substrates veiled by BN nanosheets show an outstanding reusability in the long term. As the result, the sensitivity, reproducibility and reusability of SERS substrates can be greatly improved. We also demonstrate that large BN nanosheets produced by chemical vapor deposition can be used to scale up the proposed SERS substrate for practical application.

**INTRODUCTION**

Surface enhanced Raman spectroscopy (SERS) is an ultrasensitive and nondestructive analytical technique that has a broad range of application in physics, chemistry, biology, and medicine.[1-4] Great advancements have been achieved in the field over the last decades by introducing plasmonically active metallic structures of various shapes,[5-9] and detection at single-molecule level has been realized.[10-11] Nevertheless, there remain several challenges. Conventional SERS substrates using noble metal nanostructures such as silver (Ag) or gold (Au) are not efficient in adsorption of non-thiolated aromatic molecules.[12] Such substrates also have poor reproducibility due to substrate-induced disturbances caused by metal-catalyzed side reactions, charge transfer, and photo-induced damage, etc.[13] In addition, there lacks a process that can produce effective SERS substrates at a large scale. Due to the high cost of noble metals and substrate fabrication, it is also highly desirable to achieve reusable SERS substrates. To solve these challenges, thin passivated layers of aluminum oxide ($Al_2O_3$) or silicon oxide ($SiO_2$) were deposited on SERS substrates to reduce substrate-induced disturbances and improve reproducibility and reusability.[14] However, a 1.5 nm thick $Al_2O_3$ film could weaken Raman enhancement by 75% because plasmon-induced electromagnetic fields, i.e. hot spots, decrease exponentially with distance.[15] On the other hand, 5 nm thick $Al_2O_3$ film is required to protect metal nanostructures from oxidation, as it is very difficult to produce atomically thin $Al_2O_3$ films that are pinhole-free.

Graphene can partially solve the abovementioned challenges in SERS. When used to cover plasmonic metal nanoparticles, graphene can improve the sensitivity of SERS,[16-23] reduce surface-induced disturbances,[24] and protect Ag nanoparticles from oxidation at room temperature in a short term.[25] However, graphene-based SERS substrates can hardly be reusable because graphene starts to oxidize at ~250 °C in air,[26] while reusability normally requires a heating treatment at 350 °C or higher in oxygen-containing gasses. Boron nitride (BN) nanosheets, atomically thin layers of hexagonal BN, have a similar structure to graphene but possess many different physical and chemical properties.[27-30] For example, BN nanosheets are electrically insulating with bandgaps of ~6 eV, and could sustain oxidation ~800 °C in air.[13] Therefore, they have been proposed to protect metals against oxidation and corrosion at high temperatures.[31-32] However, compared to graphene, the use of BN nanosheets for SERS has not been much investigated. Ling *et al.* studied the Raman enhancement by BN nanosheets via a chemical mechanism, but the enhancement factor was too small for practical application.[33] Lin *et al.* combined BN nanosheets with Ag nanoparticles through a solution process. The SERS substrate was reusable,[34] but its enhancement decreased 60% after the first cycle of reuse. The current authors placed Au nanoparticles on top of atomically thin BN nanosheets for SERS, and high sensitivity and good reusability were achieved.[35] Interestingly, the Raman enhancement increased with reduced thickness of BN nanosheets, suggesting superior surface adsorption capabilities of atomically thin BN.[35] More recently, Dai *et al.* used a solution process for producing porous BN microfibers decorated by Ag nanoparticles, which could efficiently capture analyte molecules for improving Raman signals and be reused after heating.[36] It is expected that there is still room to further improve the design and effectiveness of BN nanosheets for SERS.

Herein, we show for the first time that airborne organic molecules accumulated on atomically thin BN nanosheets over time. Such excellent adsorption property of BN can be valuable for SERS. For this purpose, atomically thin BN nanosheets were placed on plasmonic Au nanoparticles produced via physical routes. These SERS substrates showed dramatically improved sensitivity, reduced disturbance caused by metal nanoparticles, and outstanding stability and reusability. The design is readily scaled up using large BN nanosheets grown by chemical vapor deposition (CVD).

## EXPERIMENTAL SECTION

**Exfoliation of BN nanosheets.** BN nanosheets on 90 nm thick silicon oxide covered Si wafer ($SiO_2$/Si) were mechanically exfoliated from high-quality single crystal hBN[37] using Scotch tape. An Olympus BX51 optical microscope equipped with a DP71 camera was used to locate atomically thin nanosheets, and then a Cypher atomic force microscope (AFM) was employed to measure their thicknesses using Si cantilevers.

**BN nanosheets veiled SERS substrates.** A layer of Au film (~10 nm) was deposited on $SiO_2$/Si wafer by a Bel-Tec sputter (SCD050). The sputtering current was 40 mA, and the sputtering time was in the range of 20-40 s. BN nanosheets were mechanically exfoliated on top of the Au film following the same procedure as on $SiO_2$/Si. Subsequently, the substrates were annealed in Ar at 500 °C for 1h. For SERS tests, the substrates were immersed in rhodamine 6G (R6G, ≥95%, Fluka) or copper(II) phthalocyanine-tetrasulfonic acid tetrasodium salt (CuPc) aqueous (Milli-Q) solution for 1 h, followed by washing with water and drying in gentle Ar flow at room temperature. The CVD-grown BN nanosheet covered SERS substrate was produced following a similar procedure. In detail, CVD-grown BN (~20L thick) (Graphene Supermarket) was coated with a thin layer of poly(methyl methacrylate) (PMMA) before the copper substrate was etched and the polymer plus BN nanosheet was transferred onto $SiO_2$/Si sputtered with ~10 nm Au film. The sample was annealed in Ar for 1 h to transfer the Au film to nanoparticles.

**Characterization.** The Raman spectra were collected using a confocal Raman (Renishaw inVia) with a 514.5 nm laser. A 100x objective lens with a numerical aperture of 0.90 was used, and the laser power was ~2.5 mW. All Raman spectra were calibrated with the Si band at 520.5 $cm^{-1}$. X-ray photoelectron spectroscopy (XPS) analyses were conducted on a monolayer (1L) BN produced by CVD in an ultrahigh vacuum chamber (~$10^{-10}$ mbar) of the soft X-ray spectroscopy beamline at the Australian Synchrotron, Victoria, Australia. The excitation energy was 750 eV, the E-pass was set to 20 eV for optimum energy resolution. The excitation photon energies were calibrated by the photon energy measured on a reference Au film. The binding energies were normalized by the C–C peak at 284.5 eV. The infrared spectra represent the average of 64 scans from a Bruker Lumos infrared spectrometer in ATR mode with a resolution of 4 $cm^{-1}$.

## RESULTS AND DISCUSSION

We found that atomically thin BN nanosheets were prone to adsorb airborne organic molecules, implying their excellent surface adsorption capabilities. To more quantitatively investigate such phenomenon, the thickness change of atomically thin BN nanosheets after exposed to air for different lengths of time was measured by AFM. As-exfoliated 1L, 2L and 3L BN nanosheets had initial heights of 0.45±0.15, 0.90±0.10, and 1.10±0.10 nm, respectively, consistent with previous reports.[30,35,38] The thickness of the 1L BN increased to 0.95±0.05 nm after exposure to ambient atmosphere for 3 weeks, and further expanded to 1.25±0.25 nm in the following 9 weeks, as shown in Figure 1. BN nanosheets of 2L and 3L also experienced thickness increase in the same period: to 1.65±0.05 and 2.25±0.25 nm after 12 weeks, respectively (Figure 1d and S1). Although there has been no report on the efficient adsorption of airborne hydrocarbon on atomically thin BN nanosheets, graphene and BN nanotubes have been found to spontaneously attract airborne organics in air.[39-40] It should be noted that the exfoliated BN nanosheets were almost free of defects,[30] and hence could reflect their intrinsic surface properties.

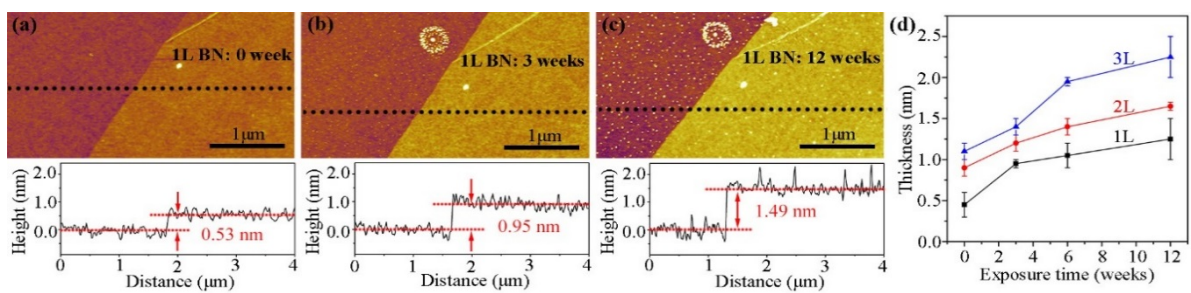

**Figure 1.** AFM images and the corresponding height traces of a 1L BN nanosheet (a) before, after (b) 3 weeks, and (c) 12 weeks of exposure to ambient atmosphere, respectively; (c) thickness changes of 1L, 2L, and 3L BN nanosheets after exposure to air.

The chemical composition of the adsorbates on atomically thin BN was analyzed using FTIR and Synchrotron-based XPS. Annealing at 450 °C in ultrahigh vacuum (~$10^{-10}$ mbar) for 4 h was used to clean the starting BN nanosheet. According to *in situ* XPS analysis, the nanosheet was almost free of carbon (arrowed in black spectrum in Figure 2a). The BN nanosheet was then exposed to air for 34 weeks, which gave rise to much stronger carbon signals in XPS due to airborne adsorbates (red spectrum in Figure 2a). According to the least fitting of the XPS spectra in the C 1s region, the adsorbates contained C–C/C–H bonds at 284.5 eV, along with C=C at 284.0 eV, C–O at 285.5 eV, C–O–C at 286.2 eV, and C=O at 288.0 eV, respectively (Figure 2b).[41-42] The FTIR data agreed well with these XPS results. The three FTIR peaks in the range of 2800 and 3000 cm$^{-1}$ can be attributed to symmetric and asymmetric stretching of methylene group (–CH$_2$–) and asymmetric stretching of –CH$_3$ group,[40,43] suggesting that the airborne hydrocarbon adsorbed on atomically thin BN nanosheets has a similar chemical composition to that on graphene.[40]

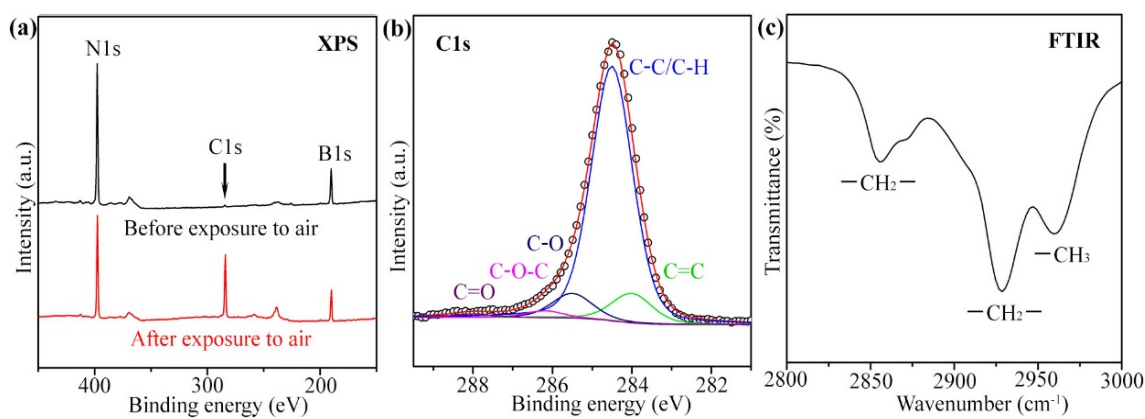

**Figure 2.** (a) XPS spectra of a 1L BN nanosheet before and after exposure to air; (b) high-resolution XPS spectrum of the C 1s peak after air exposure; (c) FTIR spectrum of the same sample after exposure to air.

The excellent adsorption property of atomically thin BN nanosheets towards organic molecules is useful for attracting analyte molecules and hence improving sensitivity in SERS. To maximize the adsorption surface, BN nanosheets were placed on top of plasmonic Au nanoparticles. It was realized by firstly sputtering an Au thin film (~10 nm) on SiO$_2$/Si wafer, then mechanically exfoliating atomically thin BN on top, and subsequently annealing the substrate to obtain Au nanoparticles. Such fabrication route is straightforward, scalable, and have good control over the size and distribution of the plasmonic metal nanoparticles.

Atomically thin BN nanosheets were located under an optical microscope, and their thickness was determined by AFM before annealing (Figure 3c). After annealing the Au film turned to nanoparticles, whose size and distribution could be controlled by sputtering time and annealing temperature, as shown in our previous study.[35] The annealing gave 1L and 2L BN more optical contrast (Figure 3a *v.s.* 3b), which greatly facilitated the following SERS measurements. As shown in Figure 3d-f, BN nanosheets of different thicknesses wrinkled to different degrees due to their different flexibilities. The surface roughness of the 1L, 2L and 13L BN nanosheets (1L/Au, 2L/Au, and 13L/Au), and bare Au particles (Au/SiO$_2$) were quantitatively characterized by height-height correlation function:

$$g(x) = \langle (h(\vec{x}) - h(\vec{x} - \vec{r}))^2 \rangle \qquad (1)$$

where $\vec{x}$ is any specific point in the image, and $\vec{r}$ is a displacement vector. The average height difference between any two points separated by a distance $r$ is described by the function $g(x)$.[44] Figure 3g shows the best fit of the function for bare Au nanoparticles, and 1L, 2L, and 13L BN nanosheets on Au nanoparticles. The root mean square (RMS) roughness of the 1L BN is ~3.5 nm, which is quite close to that of the bare Au nanoparticles (~4.0 nm), indicating that the 1L BN closely followed the profile of the underneath particles. Moreover, the correlation lengths of the bare Au nanoparticles, 1L, 2L and 13L BN covered substrates were 37, 50, 54 and 100 nm, respectively. Therefore, the roughness of BN nanosheets on Au nanoparticles increased with the decrease of nanosheet thickness. This is not surprising because atomically thin nanosheets are much more flexible and hence can deform to a much higher extent compared to thicker ones. According to SEM studies, the size distribution and separation of Au nanoparticles with and without BN coverage were quite similar (Figure 3h and Figure S3).

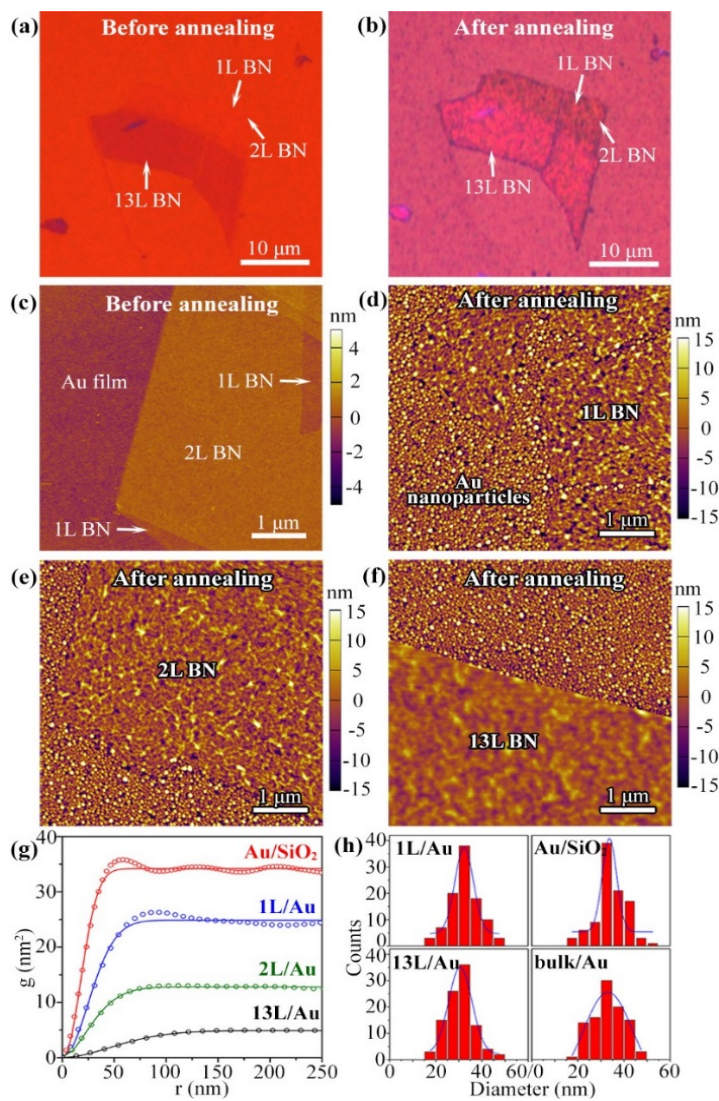

**Figure 3.** Optical microscopy images of BN nanosheets of different thicknesses on (a) Au film before annealing and (b) Au particles after annealing in Ar for 1 h; (c) AFM image of 1L and 2L BN nanosheets on Au film before annealing; AFM images of (d) 1L, (e) 2L and (f) 13L BN veiled Au particles after annealing; (g) height-height correlations of 1L, 2L, and 13L BN nanosheets on Au nanoparticles, and bare Au nanoparticles (without coverage of BN) after annealing, and the continuous lines represent the best fit according to Eq. 1; (h) size distribution of Au particles underneath 1L, 13L, and bulk BN, and bare Au nanoparticles on $SiO_2$/Si.

To test their Raman enhancement, the 1L, 13L, bulk BN covered Au substrates (abbreviated to 1L/Au, 13L/Au, bulk/Au, respectively), and bare Au particles without BN (Au/SiO$_2$) substrates were immersed in R6G aqueous solution ($10^{-7}$ M) for the same period. The Raman signals from R6G were most prominent from 1L/Au and decreased with the increase of BN thickness, e.g. 13L/Au and bulk/Au (Figure 4a). It should be emphasized that the sensitivity of 1L/Au is higher than that reported before.[34] The difference should not be caused by Au particles because their size and distribution were similar with the coverage of BN nanosheets of different thicknesses (Figure 3h). The stronger Raman signals from 1L/Au can be attributed to the better retained plasmonic hot spots. As illustrated in Figure 4b, atomically thin BN nanosheets were more flexible, and hence highly conformed to the underlying Au nanoparticles so that the analyte molecules are closer to the plasmonic hot spots; in contrast, thicker BN layers were much less deformed, and analytes were further away from the plasmon-induced electromagnetic fields which decayed exponentially with distance.[45] Molecules on Au/SiO$_2$ should have the closest distance to the hot spots, but the SERS signal was quite weak (Figure 4a). This was because Au and SiO$_2$ were inefficient in capture of R6G molecules during immersion. That is, much fewer amounts of analyte molecules were adsorbed on Au/SiO$_2$, greatly depressing the SERS signals. These were further confirmed by the measurements of the enhancement factors of different SERS substrates (see Supporting Information). In contrast, atomically thin BN has a strong adsorption capability towards aromatic molecules due to π-π interactions, and much more R6G molecules were attracted during immersion. The stronger adsorption capability of atomically thin BN nanosheets is due to conformational change, and polarity of BN should not contribute to such phenomenon. The details will be discussed elsewhere. Another advantage of the design is that BN nanosheets separate the analytes from metal nanoparticles so that substrate-induced disturbance to Raman signals is eliminated. It was not rare that Au/SiO$_2$ substrates showed extraneous Raman peaks that could not be assigned to R6G (arrowed in Figure 4c). These extraneous peaks should be due to side reactions or photo-induced damage catalyzed by Au nanoparticles.[13] In contrast, BN nanosheet veiled areas always gave "clean" Raman features of R6G (lower in Figure 4c). This is the first time that such effect has been reported on BN nanosheets as substrates for SERS. Therefore, atomically thin BN on top of plasmonic nanostructures could greatly improve SERS sensitivity and eliminate undesirable substrate-induced disturbance.

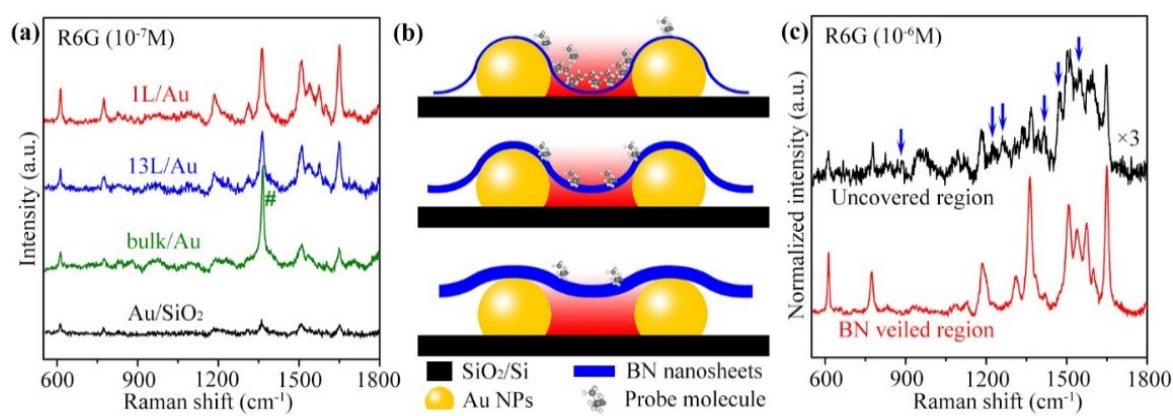

**Figure 4.** (a) SERS spectra from 1L/Au (monolayer BN on Au nanoparticles), 13L/Au (13 layer thick BN on Au nanoparticles), bulk/Au (bulk hBN on Au nanoparticles), and Au/SiO$_2$ (bare Au nanoparticles on SiO$_2$ without BN coverage) substrates after immersion in 10$^{-7}$ M R6G solution, and # represents Raman G band of bulk hBN; (b) schemes illustrating that thinner BN nanosheet covered substrate has stronger SERS enhancement. (c) SERS spectra of R6G from BN veiled region (red) and non-protected region (black), with extraneous peaks due to substrate-induced disturbance arrowed.

The BN nanosheet veiled SERS substrates have outstanding stability and reusability. As shown in our previous report,[30] BN nanosheets have excellent thermal stability: 1L BN can sustain ~800 °C in air. Kostoglou *et al.* also studied the weight loss of BN nanoplates as a function of temperature in air.[46] Hence, the adsorbed analyte molecules could be effectively removed by short time heating in air without damage of BN nanosheets. To better demonstrate how effective the regeneration process is, we used R6G and CuPc alternately as probe molecules which had different Raman features. That is, 1L/Au was first soaked in R6G solution ($10^{-6}$ M) for SERS, and then heated at 400 °C in air for 5 min to burn off R6G, and reused by immersion in CuPc solution ($10^{-6}$ M) for SERS, and so on. As shown in Figure 5a, the Raman signals of R6G and CuPc from 1L/Au did not interfere in each cycle. To show this more clearly, the highlighted region in Figure 5a is enlarged in Figure 5b. The Raman peaks at 612 and 680 cm$^{-1}$ represent C–C–C in-plane bend in R6G[47] and N–Cu stretch and outer ring stretches in CuPc,[48] respectively. The 612 cm$^{-1}$ peak was present only in R6G cycles (Cycle 0, 2, and 4), and undetectable in CuPc cycles (Cycle 1, 3 and 5), and vice versa. It means that the heating at 400 °C in air was very effective in removing these analyte molecules (black in Figure 5a and b). The spectrum after heating (black in Figure 5a) also shows that the Raman signature (*i.e.* G band) of atomically thin BN nanosheets is so weak that it does not interfere with or introduce extraneous peaks to the signals of analytes. In a control experiment, if 1L/Au was not heated before the next cycle, the signals of both R6G and CuPc were present (see Supporting Information, Figure S5). The Raman enhancement of 1L/Au did not decrease after 5 cycles (Figure 5c), as the BN nanosheet was intact from the heating treatments (see Supporting Information, Figure S6). To test its long-term stability, 1L/Au was heated at 400 °C in air for extended periods. There was no decrease of enhancement after heating for up to 1000 min (~16.7 h); the SERS sensitivity plummeted only after 2500 min (41.7 h) heating, indicating the loss of the 1L BN and its excellent adsorption towards R6G (see Supporting Information, Figure S7).

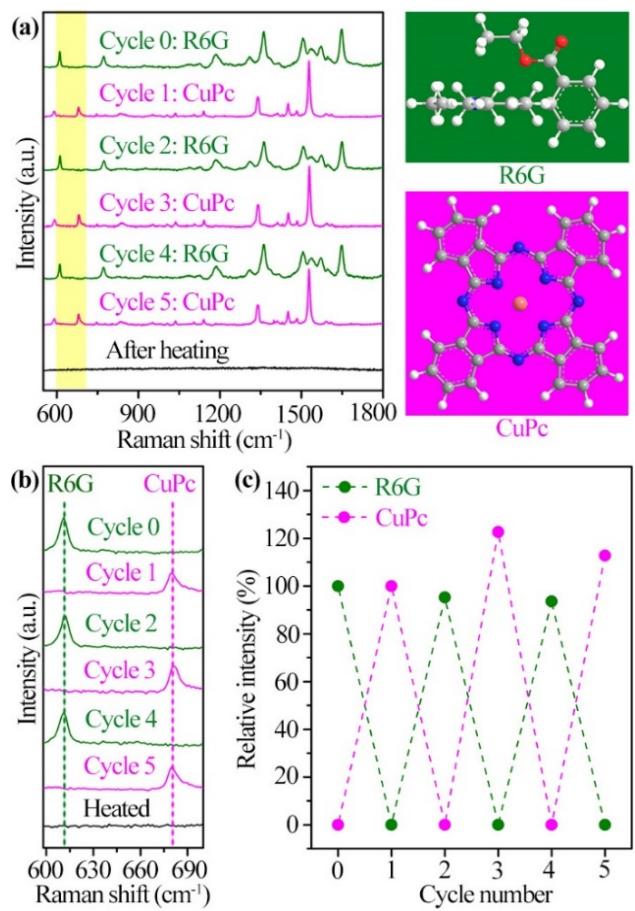

**Figure 5.** (a) Reusability tests of 1L/Au using R6G and CuPc alternately after heating at 400 °C in air for 5 min; (b) enlarged view of the highlighted region in (a); (c) relative intensity changes of 612 and 680 cm$^{-1}$ peaks after reusability cycles.

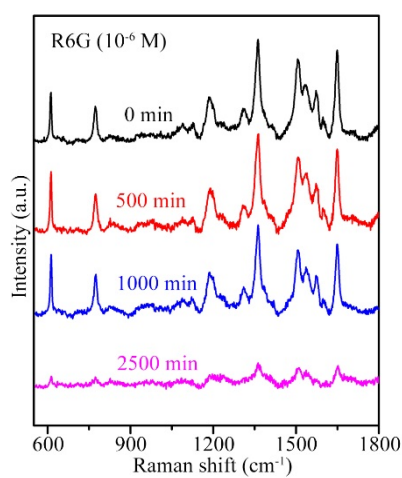

**Figure 6.** Raman spectra of R6G from 1L/Au after heat treatment at 400 °C in air for 0, 500, 1000 and 2500 min.

The proposed BN nanosheet veiled SERS substrates are easily scaled up if large BN nanosheets produced by CVD are used. Figure 7a shows a photo of a CVD-grown BN nanosheet covered Au SERS substrate (CVD-BN/Au), and the size of the BN nanosheet was ~6×4 mm. Figure 7b shows the Raman G band of the BN nanosheet. The boundary between the BN nanosheet and Au nanoparticles can be clearly seen under optical microscope (Figure 7c). Similarly, CVD-BN/Au showed stronger Raman signals than the bare Au without BN. The higher enhancement from BN nanosheet covered area can be also attributed to its higher adsorption capability. To show the homogeneity of the enhancement, Raman mapping was conducted in the squared area in Figure 7c. The mapping in Figure 7e confirms that SERS signals were much more pronounced from the BN area, and the enhancement was relatively uniform. The CVD-BN/Au substrate was also reusable (see Supporting Information, Figure S8). Therefore, BN nanosheet veiled SERS substrates have a potential in practical application.

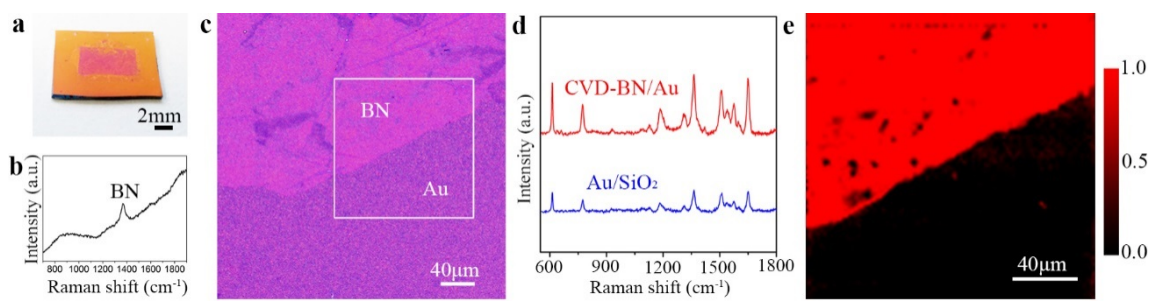

**Figure 7.** (a) Photo of a CVD-grown BN nanosheet transferred on Au substrate; (b) Raman spectra of the CVD-grown BN; (c) optical microscopy image of the CVD-grown BN on Au; (d) Raman spectra of R6G ($10^{-6}$ M) from CVD-BN/Au (upper) and Au/SiO$_2$ (lower); (e) Raman mapping of the 612 cm$^{-1}$ peak of R6G ($10^{-6}$ M) in the squared area in (c) with the lowest signal set as 0.0 and the highest as 1.0.

## CONCLUSIONS

Atomically thin BN nanosheets showed strong surface adsorption of airborne hydrocarbon and aromatic molecules. Such property makes them attractive substrates for SERS application. When placed on top of plasmonically active metal nanostructures, atomically thin BN nanosheets greatly increased the number of analyte molecules per unit area, and hence improved the sensitivity and reproducibility of SERS substrate. Furthermore, the SERS substrates veiled by BN nanosheets were highly reusable without loss of Raman enhancement after regeneration by heat treatment in air even over a long period (e.g. 400 °C for 1000 min). We also demonstrated the scale-up of the SERS substrate using CVD-grown BN nanosheet for the first time. This study contributes to an improved Raman sensitivity and promotes a wider use of SERS in various fields.

## ASSOCIATED CONTENT

**Supporting Information**. AFM images of 1-3L BN after exposure to air for different periods, SEM images of Au particles and corresponding size distribution, and Raman spectra of R6G/CuPc from 1L/Au substrate. This material is available free of charge via the Internet at http://pubs.acs.org.

## AUTHOR INFORMATION

**Corresponding Author**

*Email: luhua.li@deakin.edu.au ; ian.chen@deakin.edu.au

**Notes**

The authors declare no competing financial interests.

## ACKNOWLEDGEMENTS

L.H.Li thanks the financial support from the Australian Research Council under Discovery Early Career Researcher Award (DE160100796), and Deakin University under Alfred Deakin Research Postdoctoral Fellowship 2014, and Central Research Grants Scheme 2015. Y. Chen acknowledges the funding support from the Australian Research Council under the Discovery project. This research was partly undertaken on the soft x-ray beamline at the Australian Synchrotron, Victoria, Australia.

## REFERENCE

(1) Campion, A.; Kambhampati, P., Surface-Enhanced Raman Scattering. *Chem. Soc. Rev.* **1998,** *27* (4), 241-250.

(2) Baker, G. A.; Moore, D. S., Progress in Plasmonic Engineering of Surface-Enhanced Raman-Scattering Substrates toward Ultra-Trace Analysis. *Anal. Bioanal. Chem.* **2005,** *382* (8), 1751-1770.

(3) Haynes, C. L.; Yonzon, C. R.; Zhang, X.; Van Duyne, R. P., Surface-Enhanced Raman Sensors: Early History and the Development of Sensors for Quantitative Biowarfare Agent and Glucose Detection. *J. Raman Spectrosc.* **2005,** *36* (6-7), 471-484.

(4) Moskovits, M., Surface-Enhanced Raman Spectroscopy: A Brief Retrospective. *J. Raman Spectrosc.* **2005,** *36* (6-7), 485-496.

(5) Rycenga, M.; Xia, X.; Moran, C. H.; Zhou, F.; Qin, D.; Li, Z. Y.; Xia, Y., Generation of Hot Spots with Silver Nanocubes for Single-Molecule Detection by Surface-Enhanced Raman Scattering. *Angew. Chem. Int. Ed.* **2011,** *123* (24), 5587-5591.

(6) Li, L.; Hutter, T.; Steiner, U.; Mahajan, S., Single Molecule SERS and Detection of Biomolecules with a Single Gold Nanoparticle on a Mirror Junction. *Analyst* **2013,** *138* (16), 4574-4578.

(7) Lee, J.; Hua, B.; Park, S.; Ha, M.; Lee, Y.; Fan, Z.; Ko, H., Tailoring Surface Plasmons of High-Density Gold Nanostar Assemblies on Metal Films for Surface-Enhanced Raman Spectroscopy. *Nanoscale* **2014,** *6* (1), 616-623.

(8) Tao, A.; Kim, F.; Hess, C.; Goldberger, J.; He, R.; Sun, Y.; Xia, Y.; Yang, P., Langmuir-Blodgett Silver Nanowire Monolayers for Molecular Sensing Using Surface-Enhanced Raman Spectroscopy. *Nano Lett.* **2003,** *3* (9), 1229-1233.

(9) Wang, P.; Liang, O.; Zhang, W.; Schroeder, T.; Xie, Y. H., Ultra-Sensitive Graphene-Plasmonic Hybrid Platform for Label-Free Detection. *Adv. Mater.* **2013,** *25* (35), 4918-4924.

(10) Kneipp, K.; Wang, Y.; Kneipp, H.; Perelman, L. T.; Itzkan, I.; Dasari, R. R.; Feld, M. S., Single Molecule Detection Using Surface-Enhanced Raman Scattering (SERS). *Phys. Rev. Lett.* **1997,** *78* (9), 1667-1670.

(11) Nie, S.; Emory, S. R., Probing Single Molecules and Single Nanoparticles by Surface-Enhanced Raman Scattering. *Science* **1997,** *275* (5303), 1102-1106.

(12) Zheng, J.; Li, X.; Gu, R.; Lu, T., Comparison of the Surface Properties of the Assembled Silver Nanoparticle Electrode and Roughened Silver Electrode. *J. Phys. Chem. B* **2002,** *106* (5), 1019-1023.

(13) Le Ru, E.; Etchegoin, P., *Principles of Surface-Enhanced Raman Spectroscopy: And Related Plasmonic Effects*. Elsevier: **2008**.

(14) Li, J. F.; Huang, Y. F.; Ding, Y.; Yang, Z. L.; Li, S. B.; Zhou, X. S.; Fan, F. R.; Zhang, W.; Zhou, Z. Y.; Ren, B., Shell-Isolated Nanoparticle-Enhanced Raman Spectroscopy. *Nature* **2010,** *464* (7287), 392-395.

(15) Dieringer, J. A.; McFarland, A. D.; Shah, N. C.; Stuart, D. A.; Whitney, A. V.; Yonzon, C. R.; Young, M. A.; Zhang, X. Y.; Van Duyne, R. P., Surface Enhanced Raman Spectroscopy: New Materials, Concepts, Characterization Tools, and Applications. *Faraday Discuss* **2006,** *132*, 9-26.

(16) Xu, W.; Mao, N.; Zhang, J., Graphene: A Platform for Surface-Enhanced Raman Spectroscopy. *Small* **2013,** *9* (8), 1206-1224.

(17) Xu, S. C.; Wang, J. H.; Zou, Y.; Liu, H. P.; Wang, G. Y.; Zhang, X. M.; Jiang, S. Z.; Li, Z.; Cao, D. Y.; Tang, R. X., High Performance SERS Active Substrates Fabricated by Directly Growing Graphene on Ag Nanoparticles. *RSC Adv.* **2015,** *5* (110), 90457-90465.

(18) Fan, W.; Lee, Y. H.; Pedireddy, S.; Zhang, Q.; Liu, T. X.; Ling, X. Y., Graphene Oxide and Shape-Controlled Silver Nanoparticle Hybrids for Ultrasensitive Single-Particle Surface-Enhanced Raman Scattering (SERS) Sensing. *Nanoscale* **2014,** *6* (9), 4843-4851.

(19) Kang, L. L.; Chu, J. Y.; Zhao, H. T.; Xu, P.; Sun, M. T., Recent Progress in the Applications of Graphene in Surface-Enhanced Raman Scattering and Plasmon-Induced Catalytic Reactions. *J. Mater. Chem. C* **2015,** *3* (35), 9024-9037.

(20) Liang, X.; Liang, B. L.; Pan, Z. H.; Lang, X. F.; Zhang, Y. G.; Wang, G. S.; Yin, P. G.; Guo, L., Tuning Plasmonic and Chemical Enhancement for SERS Detection on Graphene-Based Au Hybrids. *Nanoscale* **2015,** *7* (47), 20188-20196.

(21) Zhao, Y.; Zhu, Y. W., Graphene-Based Hybrid Films for Plasmonic Sensing. *Nanoscale* **2015,** *7* (35), 14561-14576.

ACS Applied Materials & Interfaces 8, 15630, 2016
(DOI: 10.1021/acsami.6b04320)

(22) Osvath, Z.; Deak, A.; Kertesz, K.; Molnar, G.; Vertesy, G.; Zambo, D.; Hwang, C.; Biro, L. P., The Structure and Properties of Graphene on Gold Nanoparticles. *Nanoscale* **2015,** *7* (12), 5503-5509.

(23) Zhang, X.; Shi, C. S.; Liu, E. Z.; Li, J. J.; Zhao, N. Q.; He, C. N., Nitrogen-Doped Graphene Network Supported Copper Nanoparticles Encapsulated with Graphene Shells for Surface-Enhanced Raman Scattering. *Nanoscale* **2015,** *7* (40), 17079-17087.

(24) Xu, W.; Ling, X.; Xiao, J.; Dresselhaus, M. S.; Kong, J.; Xu, H.; Liu, Z.; Zhang, J., Surface Enhanced Raman Spectroscopy on a Flat Graphene Surface. *Proc. Natl. Acad. Sci. U. S. A.* **2012,** *109* (24), 9281-9286.

(25) Li, X.; Li, J.; Zhou, X.; Ma, Y.; Zheng, Z.; Duan, X.; Qu, Y., Silver Nanoparticles Protected by Monolayer Graphene as a Stabilized Substrate for Surface Enhanced Raman Spectroscopy. *Carbon* **2014,** *66*, 713-719.

(26) Liu, L.; Ryu, S. M.; Tomasik, M. R.; Stolyarova, E.; Jung, N.; Hybertsen, M. S.; Steigerwald, M. L.; Brus, L. E.; Flynn, G. W., Graphene Oxidation: Thickness-Dependent Etching and Strong Chemical Doping. *Nano Lett.* **2008,** *8* (7), 1965-1970.

(27) Li, L. H.; Chen, Y., Atomically Thin Boron Nitride: Unique Properties and Applications. *Adv. Funct. Mater.* **2016**, *26*, 2594-2608.

(28) Lindsay, L.; Broido, D., Enhanced Thermal Conductivity and Isotope Effect in Single-Layer Hexagonal Boron Nitride. *Phys. Rev. B* **2011,** *84* (15), 155421.

(29) Zhi, C.; Bando, Y.; Tang, C.; Kuwahara, H.; Golberg, D., Large-Scale Fabrication of Boron Nitride Nanosheets and Their Utilization in Polymeric Composites with Improved Thermal and Mechanical Properties. *Adv. Mater.* **2009,** *21* (28), 2889-2893.

(30) Li, L. H.; Cervenka, J.; Watanabe, K.; Taniguchi, T.; Chen, Y., Strong Oxidation Resistance of Atomically Thin Boron Nitride Nanosheets. *ACS Nano* **2014,** *8* (2), 1457-1462.

(31) Liu, Z.; Gong, Y.; Zhou, W.; Ma, L.; Yu, J.; Idrobo, J. C.; Jung, J.; MacDonald, A. H.; Vajtai, R.; Lou, J., Ultrathin High-Temperature Oxidation-Resistant Coatings of Hexagonal Boron Nitride. *Nat. Commun.* **2013,** *4*, 2541.

(32) Li, L. H.; Xing, T.; Chen, Y.; Jones, R., Boron Nitride Nanosheets for Metal Protection. *Adv. Mater. Interfaces* **2014,** *1* (8), 1300132.

(33) Ling, X.; Fang, W.; Lee, Y.-H.; Araujo, P. T.; Zhang, X.; Rodriguez-Nieva, J. F.; Lin, Y.; Zhang, J.; Kong, J.; Dresselhaus, M. S., Raman Enhancement Effect on Two-Dimensional Layered Materials: Graphene, h-BN and $MoS_2$. *Nano Lett.* **2014,** *14* (6), 3033-3040.

(34) Lin, Y.; Bunker, C. E.; Fernando, K. S.; Connell, J. W., Aqueously Dispersed Silver Nanoparticle-Decorated Boron Nitride Nanosheets for Reusable, Thermal Oxidation-Resistant Surface Enhanced Raman Spectroscopy (SERS) Devices. *ACS Appl. Mater. Interfaces.* **2012,** *4* (2), 1110-1117.

(35) Cai, Q.; Li, L. H.; Yu, Y.; Liu, Y.; Huang, S.; Chen, Y.; Watanabe, K.; Taniguchi, T., Boron Nitride Nanosheets as Improved and Reusable Substrates for Gold Nanoparticles Enabled Surface Enhanced Raman Spectroscopy. *Phys. Chem. Chem. Phys.* **2015,** *17* (12), 7761-7766.

(36) Dai, P.; Xue, Y.; Wang, X.; Weng, Q.; Zhang, C.; Jiang, X.; Tang, D.; Wang, X.; Kawamoto, N.; Ide, Y., Pollutant Capturing SERS Substrate: Porous Boron Nitride Microfibers with Uniform Silver Nanoparticle Decoration. *Nanoscale* **2015,** *7* (45), 18992-18997.

(37) Taniguchi, T.; Watanabe, K., Synthesis of High-Purity Boron Nitride Single Crystals under High Pressure by Using Ba–BN Solvent. *J. Cryst. Growth* **2007,** *303* (2), 525-529.

(38) Li, L. H.; Santos, E. J.; Xing, T.; Cappelluti, E.; Roldán, R.; Chen, Y.; Watanabe, K.; Taniguchi, T., Dielectric Screening in Atomically Thin Boron Nitride Nanosheets. *Nano Lett.* **2014,** *15* (1), 218-223.

(39) Boinovich, L. B.; Emelyanenko, A. M.; Pashinin, A. S.; Lee, C. H.; Drelich, J.; Yap, Y. K., Origins of Thermodynamically Stable Superhydrophobicity of Boron Nitride Nanotubes Coatings. *Langmuir* **2011,** *28* (2), 1206-1216.

(40) Li, Z.; Wang, Y.; Kozbial, A.; Shenoy, G.; Zhou, F.; McGinley, R.; Ireland, P.; Morganstein, B.; Kunkel, A.; Surwade, S. P., Effect of Airborne Contaminants on the Wettability of Supported Graphene and Graphite. *Nat. Mater.* **2013,** *12* (10), 925-931.

(41) Tien, H.-W.; Huang, Y.-L.; Yang, S.-Y.; Hsiao, S.-T.; Liao, W.-H.; Li, H.-M.; Wang, Y.-S.; Wang, J.-Y.; Ma, C.-C. M., Preparation of Transparent, Conductive Films by Graphene Nanosheet Deposition on Hydrophilic or Hydrophobic Surfaces through Control of the Ph Value. *J. Mater. Chem.* **2012,** *22* (6), 2545-2552.

(42) Qiu, B.; Zhou, Y.; Ma, Y.; Yang, X.; Sheng, W.; Xing, M.; Zhang, J., Facile Synthesis of the $Ti^{3+}$ Self-Doped $TiO_2$-Graphene Nanosheet Composites with Enhanced Photocatalysis. *Sci. Rep.* **2015,** *5*, 8591.

(43) Shinozaki, A.; Arima, K.; Morita, M.; Kojima, I.; Azuma, Y., Ftir-Atr Evaluation of Organic Contaminant Cleaning Methods for $SiO_2$ Surfaces. *Anal. Sci.* **2003,** *19* (11), 1557-1559.

(44) Gredig, T.; Silverstein, E. A.; Byrne, M. P. Height-Height Correlation Function to Determine Grain Size in Iron Phthalocyanine Thin Films, *J. Phys. Conf. Ser.* **2013**, *417* (1), 012069.

(45) Le Ru, E.; Etchegoin, P.; Meyer, M., Enhancement Factor Distribution around a Single Surface-Enhanced Raman Scattering Hot Spot and Its Relation to Single Molecule Detection. *J. Chem. Phys.* **2006,** *125* (20), 204701.

(46) Kostoglou, N.; Polychronopoulou, K.; Rebholz, C., Thermal and Chemical Stability of Hexagonal Boron Nitride (h-BN) Nanoplatelets. *Vacuum* **2015,** *112*, 42-45.

(47) Zhai, W.-L.; Li, D.-W.; Qu, L.-L.; Fossey, J. S.; Long, Y.-T., Multiple Depositions of Ag Nanoparticles on Chemically Modified Agarose Films for Surface-Enhanced Raman Spectroscopy. *Nanoscale* **2012,** *4* (1), 137-142.

(48) Jiang, N.; Foley, E.; Klingsporn, J.; Sonntag, M.; Valley, N.; Dieringer, J.; Seideman, T.; Schatz, G.; Hersam, M.; Van Duyne, R., Observation of Multiple Vibrational Modes in Ultrahigh Vacuum Tip-Enhanced Raman Spectroscopy Combined with Molecular-Resolution Scanning Tunneling Microscopy. *Nano Lett.* **2012,** *12* (10), 5061-5067.

**TOC**

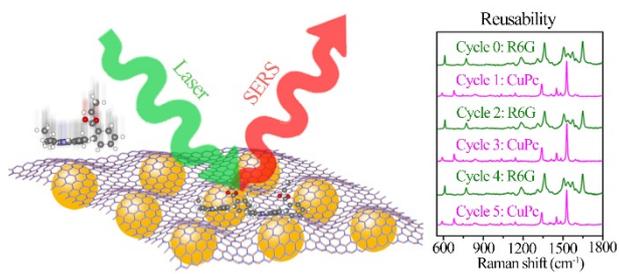